\documentclass[useAMS,usenatbib]{mn2e}

\usepackage{epsf}

\usepackage{color}
\definecolor{Blue}{cmyk}{1.0,1.0,0.0,0.0}
\definecolor{Green}{cmyk}{1.0,0.0,1.0,0.0}
\definecolor{Orange}{cmyk}{0.0,0.5,1.0,0.0}
\definecolor{Red}{cmyk}{0.0,1.0,1.0,0.0}

\title[Measuring spin of a supermassive black hole at the Galactic
  Centre]{Measuring spin of a supermassive black hole at the Galactic
  Centre -- Implications for a unique spin}
\author[Y. Kato, M. Miyoshi, R. Takahashi, H. Negoro, and
  R. Matsumoto]{Y. Kato$^{1}$\thanks{E-mail:
kato.yoshiaki@isas.jaxa.jp (YK)}, M. Miyoshi$^{2}$,
  R. Takahashi$^{3}$, H. Negoro$^{4}$, R. Matsumoto$^{5}$\\
$^{1}$Japan Aerospace Exploration Agency (JAXA), 3-1-1 Yoshinodai,
  Sagamihara, Kanagawa 229-8510, Japan \\
$^{2}$National Astronomical Observatory of Japan, Mitaka, Tokyo
  181-8588, Japan \\
$^{3}$The Institute of Physical and Chemical Research (RIKEN), Wako,
  Saitama 351-0198, Japan \\
$^{4}$Department of Physics, College of Science and Technology, Nihon
  University, 1-8 Kanda-Surugadai, Chiyoda-ku, Tokyo 101-8308, Japan\\
$^{5}$Department of Physics, Graduate School of Science, Chiba
  University, 1-33 Yayoi-Cho, Inage-Ku, Chiba 263-8522, Japan}
\begin{document}

\date{Submitted 2009 June 30}

\pagerange{\pageref{firstpage}--\pageref{lastpage}} \pubyear{2002}

\maketitle

\label{firstpage}

\begin{abstract}
We determine the spin of a supermassive black hole in the context
of discseismology by comparing newly detected quasi-periodic
oscillations (QPOs) of radio emission in the Galactic centre,
Sagittarius A* (Sgr A*), as well as infrared and X-ray emissions with
those of the Galactic black holes.
We find that the spin parameters of black holes in Sgr A* and in
Galactic X-ray sources have a unique value of $\approx 0.44$ which is
smaller than the generally accepted value for supermassive black holes,
suggesting evidence for the angular momentum extraction of black holes
during the growth of supermassive black holes.
Our results demonstrate that the spin parameter approaches the
equilibrium value where spin-up via accretion is balanced by spin-down
via the Blandford-Znajek mechanism regardless of its initial spin.
We anticipate that measuring the spin of black holes by using QPOs
will open a new window for exploring the evolution of black holes in
the Universe.
\end{abstract}


\section{Introduction}
The Galactic centre, Sagittarius A* (Sgr A*), is a compact source of
radio, infrared, and X-ray emissions having variability in the range
of a few tens of minutes to hours (Baganoff et al. 2001; Genzel et
al. 2003; Yusef-Zadeh et al. 2006).
These emissions seem to originate from a hot and low-density
accreting gas plunging into a supermassive black hole (Yuan et
al. 2004; Kato et al. 2009).
A precise measurement of its mass and spin is a long-standing issue
for astrophysics to investigate the mechanism of energy extraction
from spinning black holes for astrophysical jet production as well as
the evolution of supermassive black holes along the cosmic hitory
(Bardeen 1970; Blandford~\& Znajek 1977; Wilson~\& Cobert 1995).
Although the mass of Sgr A* has been constrained by using the stellar
orbit method, a precise measurement of its spin for the best-estimated
mass has been poorly conducted.

Recently, Miyoshi and colleagues have detected multiple quasi-periodic
oscillations (QPOs) of radio emissions in Sgr A* (Miyoshi et al. in
prep.), whose periods are close to the Keplerian period at the
innermost stable circular orbit of a supermassive black hole with mass
$4\times 10^{6} M_{\odot}$. Because of the excellent spatial
resolution of the Very Long Baseline Array (VLBA), the quasi-periodic
radio emission certainly originates from within the central sub-mas
scale, approximately 100 $r_{\rm g}$ around the central black hole at
a distance of 7.6 kpc, where $r_{\rm
  g}=GM/c^{2}=0.01\left(M/10^{6}M_{\odot}\right)$ AU is the
gravitational radius ($G$, $M$, and $c$ are the gravitational
constant, the mass of black hole, and the speed of light,
respectively).  This is the first time that such multiple QPOs have
been identified in the vicinity of a supermassive black hole. The
spatial pattern of emission regions cannot be explained by the
Keplerian rotation of a single emitting body at a given radius.  

Four simultaneous QPOs (16.8, 22.2, 31.4, and 56.4 min) are detected
and the first three periods are identical to QPOs in the near infrared
and X-ray observations during different observation epochs (see Table
1).  Three identical periods in the different wavelength are stable at
least for several years and the frequency ratio of last two periods is
close to 3:2.  Such a stable double peak QPO is a well-known feature
for high-frequency QPOs (HF-QPOs) in Galactic X-ray
sources (Remillard~\& McClintock 2006).  The multiple periodicity and
their coincidence between the different wavelengths, and also the
different observation epochs, indicates that the origin of QPOs in the
Galactic centre is closely related to the dynamics of an accretion
disc feeding the black hole.  Therefore we measure the spin parameter
of a black hole in Sgr A* by using the period of QPOs based on
discseismology (e.g., Nowak~\& Wagoner 1993).

\begin{table*}
\begin{center}
\caption{QPOs detected in Sgr A*.}
{\small 
\begin{tabular}{lcccl}\hline
Obs. epoch (UT) & Obs. band & Period (min) & Ref. \# \\\hline
2003/06/15 - 16 & K-band & $16.8\pm 2, 28.0$ & Genzel et al. 2003 \\
2004/09 & $1.60, 1.87, 1.90$ ($\mu$m) & $33\pm2$ & Yusef-Zadeh et al. 2006 \\
2002/10, 2004.08 & $2 - 10$ (keV) & $22.2$ & B\'{e}langer et al. 2006 \\
2007/04/04 & L-band & $22.6$ & Hamaus et al. 2009 \\
2007/07/22 & L-band & $45.4$ & Hamaus et al. 2009 \\
2004/03/08 09:30 - 16:30 & $43$ (GHz) & $16.8\pm 1.4, 22.2\pm 1.4,
31.2\pm 1.5, 56.4\pm 6$ & Miyoshi et al. in prep. \\\hline
\end{tabular}}
\end{center}
\end{table*}

\section{Method and Model}
One promising mechanism of generating multiple QPOs is a global disc
oscillation excited by the resonance between geodesic modes of
the disc (the so-called {\it resonant disc oscillation}
model: Abramowicz~\& Klu{\'z}niak 2001; Kato~\& Fukue 2006; Kato et
al. 2008).  The resonant frequency is the combination among geodesic
frequencies at the radius where the resonance occurs.  When the
resonance condition is specified, both the resonant frequency and the
resonant radius is determined uniquely in terms of the black hole mass
$M$ and the spin parameter $a_{*}\equiv Jc/GM^{2}$ where $J$ is the
angular momentum of the black hole.  Therefore the metric of the black
hole can be   constrained by the frequency of the QPOs.

Resonance may occur at a radius where the frequency ratio of the
geodesic modes is a ratio of small integers and resonant response can
either spontaneously grow or damp the oscillation itself
(Abramowicz~\& Klu{\'z}niak 2001).  One of the most prominent
resonances is a mode-coupling between acoustic waves and
non-axisymmetric modes such as a warp in the disc, the so-called
wave-warp resonance (Kato~\& Fukue 2006; Kato et al. 2008).  For
example, this resonance is excited at a radius $r_{\rm res}$ where
$\Omega_{\rm K}=2\kappa$.  Here $\Omega_{\rm K}$ and $\kappa$ are the
Kepler frequency and the epicyclic frequency, respectively (see Fig. 1
of Kato~\& Fukue 2006 for the relation between $a_{*}$ and $r_{\rm
  res}$).  $\Omega_{\rm K}$ and $\kappa$ at the resonant radius
$\tilde{r}_{\rm res}=r_{\rm res}/r_{\rm g}$ measured at infinity are
expressed as
\begin{equation}
\Omega_{\rm
  K}=\sqrt{\frac{GM}{r_{\rm res}^{3}}}\left[1+\frac{a_{*}}{\tilde{r}_{\rm res}^{3/2}}\right]^{-1}
\end{equation}
and
\begin{equation}
\kappa=\sqrt{\frac{GM}{r_{\rm res}^{3}}}\frac{\sqrt{1 - 3/\tilde{r}_{\rm res} +
    8a_{*}\left(2\tilde{r}_{\rm res}\right)^{-3/2} -
    3a_{*}^{2}\left(2\tilde{r}_{\rm res}\right)^{-2}}}{1 +
  a_{*}\left(2\tilde{r}_{\rm res}\right)^{-3/2}}
\end{equation}
as derived by Okazaki et al. (1987).  The resulting frequencies
of QPOs are $m\Omega_{K}\pm\kappa$ and $m\Omega_{\rm K}$ where $m$ is
the azimuthal mode number, and some lower mode oscillations related to
such resonances are reported by numerical studies (Kato 2004).

\section{Results}
\subsection{Unified model of QPOs}
Figure 1 shows the period of the observed QPO overlayed with lower
mode ($m=1,2$) resonant periods related to the wave-warp resonance as
a function of the black hole mass ranging from a stellar mass black
hole to a supermassive black hole (skipping over the intermediate mass
region).  QPOs in the Galactic centre are selected with regard to the
multiple detection among different wavelengths (Table 1).  We found
that such QPOs in Sgr A* detected at identical frequencies are
consistent with a mass-period relation for the spin parameter
$a_{*}\sim 0.4$  (see Fig. 1b).  At the same time, HF-QPOs in the
Galactic X-ray sources agree well with the resonant periods for the
same spin parameter within the error of the estimated mass (Fig. 1a).
Therefore we identify the three identical periods (16.8, 22.2, and
31.4 min) with resonant modes $2\Omega_{\rm K}$, $\Omega_{\rm K} +
\kappa$, and $\Omega_{\rm K}$, respectively.

\begin{figure}
  \centerline{\epsfxsize=\hsize \epsfbox{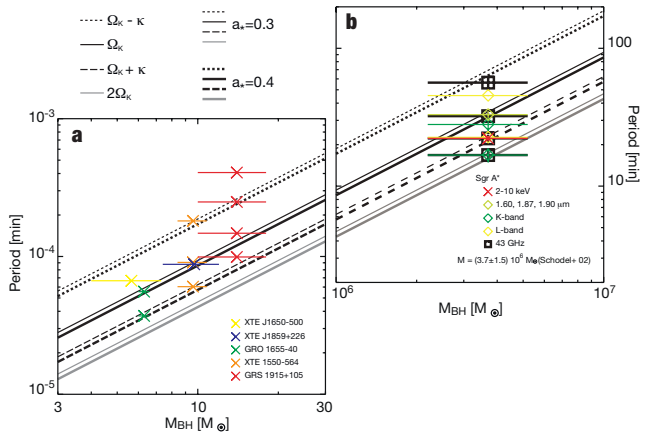}}
  \caption{Observed QPO periods as a function of black hole mass.  (a)
    HF-QPO periods of different sources are shown as crosses with
    horizontal bars indicating the range of black hole mass
    (Abramowicz~\& Klu{\'z}niak 2001; Homan et al. 2003; Orosz et
    al. 2004; Remillard~\& McClintock 2006; Kato et al. 2008).  (b)
    QPO periods of Sgr A* in different energy bands are
    shown (see Table 1).  The black hole mass is assumed to be
    $\left(3.7\pm 1.5\right)\times 10^{6}M_{\odot}$ (Sch{\"o}del et
    al. 2002).   Resonant oscillations for $m=1$ and $2$ are shown as
    solid ($\Omega_{\rm K}$), dashed ($\Omega_{\rm K} + \kappa$),
    dotted ($\Omega_{\rm K} - \kappa$), and gray solid ($2\Omega_{\rm
      K}$) lines.  Note that $2\Omega_{\rm K} - \kappa = \Omega_{\rm
      K} + \kappa$ and $2\Omega_{\rm K} + \kappa$ are omitted for
    simplicity.  Thin and thick lines indicate the periods for the
    spin parameter $a_{*}=0.3$ and $0.4$, respectively.}
  \label{fig1:eps}
\end{figure}

\subsection{Unique spin parameter}
Now we can determine the spin parameter of black holes by using the
periods of QPOs corresponding to $\Omega_{\rm K}$.  For instance,
$31.4$ min is used for  Sgr A* and periods of lower HF-QPOs are used
for the Galactic X-ray sources. Note that the frequency of
single peak HF-QPOs are treated as $\Omega_{\rm K}$.  In order to
constrain the resultant spin parameter, the estimated mass of a
supermassive black hole in Sgr A* is taken from recent
measurements (Sch{\"o}del et al. 2002; Ghez et al. 2008; Gillessen et
al. 2009).  Figure 2 shows spin parameters of all samples evaluated by
using the discseismic measurement.  All spin parameters are relatively
small ($\le 0.7$) in comparison with the equilibrium value of spinning
black holes ($\approx 0.95$) predicted by a numerical study (Gammie et
al. 2004).  When all samples are fitted by using a linear relation as
a function of the black hole mass, the spin parameter becomes larger
than $1$ for black holes with $M\geq 10^{7}\,M_{\odot}$.  Instead of a
linear relation, we obtain a best-fit unique spin parameter
$a_{*}=0.44\pm 0.08$ , which is depicted by a gray shaded region, for
$1\sigma$ uncertainty by linear least square fitting.

\begin{figure}
  \centerline{\epsfxsize=\hsize \epsfbox{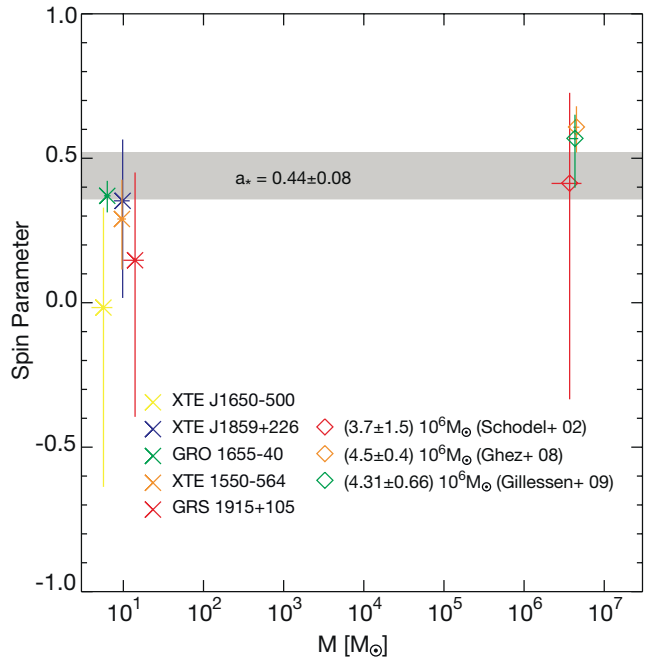}}
  \caption{Spin parameters measured by discseismic method.  Crosses
  indicate spins for the Galactic X-ray sources whereas diamonds
  indicate those for the Galactic centre in terms of black hole masses
  measured by the stellar orbits method (Sch{\"o}del et al. 2002; Ghez
  et al. 2008; Gillessen et al. 2009).  A gray shaded region indicate
  the best-fit spin parameter $a_{*}=0.44\pm 0.08$ for $1\sigma$
  uncertainty.}
  \label{fig2:eps}
\end{figure}

\subsection{Evolution of BH spin and mass}
Next, we should ask why black holes have a unique spin parameter in
spite of the fact that their age as well as mass accretion history may
vary in general.  Actually, our results contradict recent studies that
predict extremely spinning black holes (Shapiro 2005; Volonteri et
al. 2005).  In order to test the feasibility of such a small unique
spin parameter, we have to study the spin-up process by mass accretion
and the spin-down process by the energy extraction as a result of the
Blandford-Znajek mechanism, simultaneously.

Figure 3 represents the equilibrium value of spin and also the time
evolution of black holes surrounded by a relativistic standard
accretion disc (Novikov~\& Thorne 1973; Page~\& Thorne 1974; see also
Kato et al. 2008), assuming given disc parameters such
as the viscosity parameter $\alpha$ (Shakura~\& Sunyaev 1973), the
magnetized parameter $\beta$, the ratio of the gas pressure to the
magnetic pressure, and the mass accretion rate
$\dot{m}=\dot{M}/\dot{M}_{\rm EDD}$ normalized by the Eddington mass
accretion rates $\dot{M}_{\rm EDD}=4\pi G M/c\kappa_{\rm es}$
  where $\kappa_{\rm es}$ is the electron scattering opacity (e.g.,
  Kato et al. 2008). In general, these parameters are not independent
because magnetohydrodynamic (MHD) turbulence in the disc is thought to
be the source of viscosity and their values can only be examined
numerically.  For instance, we employ $\alpha=0.01$ on the basis of
three-dimensional MHD simulations showing the total stress corresponds
to $\alpha\approx 0.02 - 0.06$ (Hawley 2000; Machida et al. 2000) for
$\dot{m}\ll 1$ and $\alpha\approx 0.01$ for $\dot{m}\sim 1$ (Hirose et
al. 2006).
Recent MHD simulations also exhibit the natural emergence of
large-scale magnetic fields (the so-called magnetic tower) at the
inner region of an accretion disc (Kato et al. 2004).  The formation
of a magnetic tower is key to the extraction of the energy and angular
momentum of a spinning black hole by the Blandford-Znajek mechanism
and it has been suggested that the necessary condition for the
  energy and angular momentum extraction at the innermost region of an
  accretion disk is $\beta\approx 1$ (McKinney~\& Gammie 2004).

The equations we solved in this study are the followings:
\begin{equation}
\label{basic_eq1}
\frac{d\ln{M}}{dt}=\frac{\dot{M}}{M}e_{\rm in} - \frac{{\cal P}}{Mc^{2}}
\end{equation}
\begin{equation}
\label{basic_eq2}
\frac{dJ}{dt}=\dot{M}l_{\rm in} - \frac{{\cal P}}{\Omega_{\rm F}}
\end{equation}
where $\dot{M}$, $e_{\rm in}$, and $l_{\rm in}$ are the mass accretion
rate, the specific energy and the specific angular momentum at the
inner edge of the accretion disc, respectively.  The electromagnetic
power loss ${\cal P}$ from the black hole is assumed to be that of the
Blandford-Znajek mechanism:
\begin{equation}
{\cal P}={\cal P}_{\rm BZ}\simeq\frac{1}{8}\frac{B_{\perp}^{2}{r}_{\rm H}^{4}}{c}\Omega_{\rm F}\left(\Omega_{\rm H} - \Omega_{\rm F}\right)
\end{equation}
where ${r}_{\rm H}$ is the radius of the event horizon and
$\Omega_{\rm F}$ and $\Omega_{\rm H}$ are the angular velocity of the
magnetic fields permeating the horizon and the angular velocity of the
black hole, respectively (see Moderski~\& Sikora 1996; Beskin et
al. 2003).  The strength of magnetic fields $B_{\perp}$ permeating the
event horizon is assumed to be regulated by the pressure of accretion
disc $p_{\rm disc}$ so that $B_{\perp}^{2}=8\pi p_{\rm disc}/\beta$.
Note that the electromagnetic power loss is not negligible when
$\beta$ is less than the order of the unity.

The relativistic standard accretion disc model provides a complete
set of equations for describing the pressure of accretion disc at the
given radius as a function of the viscosity parameter $\alpha$, the
black hole mass $m=M/M_{\odot}$, the spin parameter $a_{*}$, and the
mass accretion rate $\dot{m}$.  For a given $\dot{m}$, the radiation
pressure dominated region appears within the radius:
\begin{eqnarray}
\tilde{r}_{\rm b}&=& r_{\rm b}/r_{\rm g} \cr
&=& 36\alpha^{2/21}m^{2/21}\dot{m}^{16/21}{\cal B}^{-16/21}{\cal D}^{2/21}{\cal H}^{-10/21}{\cal Q}^{16/21}
\end{eqnarray}
where ${\cal B}$, ${\cal D}$, ${\cal H}$, and ${\cal Q}$ are the
general relativistic correction factors (Page \&  Thorne 1974).  To
summarize, the pressure of the accretion disc can be described as
follows:
\begin{equation}
p_{\rm disc} =\left\{\begin{array}{ll}
p_{\rm rad} & {\tilde r}\leq\tilde{r}_{\rm b},\\
p_{\rm gas} & {\tilde r} > \tilde{r}_{\rm b},
\end{array}\right.
\end{equation}
and
\begin{equation}
p_{\rm rad}=1.4\times 10^{16}\left(\alpha m\right)^{-1}{\cal
  R}_{1}\hspace{2mm}{\rm dyne}\,{\rm cm^{-2}},
\end{equation}
\begin{equation}
p_{\rm gas}=3.0\times 10^{17}\left(\alpha m\right)^{-9/10}\dot{m}^{4/5}{\cal
  R}_{2}\hspace{2mm}{\rm dyne}\,{\rm cm^{-2}},
\end{equation}
where ${\cal R}_{1}=\tilde{r}^{-3/2}{\cal B}^{-2}{\cal D}^{-1}{\cal
  C}$ and ${\cal R}_{2}=\tilde{r}^{-51/20}{\cal B}^{-14/5}{\cal
  D}^{-9/10}{\cal C}{\cal H}^{-1/2}{\cal Q}^{4/5}$ are the radial
dependence including the general relativistic correction factors at
the Boyer-Lindquist coordinated radius $\tilde{r}=c^{2}r/GM$.  The
radius for evaluating the strength of magnetic field is asssumed to be
$\tilde{r}_{0} = 1.3\tilde{r}_{\rm ms}$ where $\tilde{r}_{\rm ms}$ is
the marginally stable circular orbit (Bardeen et al. 1972):
\begin{equation}
\tilde{r}_{\rm ms}=3 + z_{2} - \left\{\left(3-z_{1}\right)\left(3+z_{1}+2z_{2}\right)\right\}^{1/2},
\end{equation}
where
\begin{equation}
z_{1}=1+\left(1-a_{*}^{2}\right)^{1/3}\left[\left(1+a_{*}\right)^{1/3}+\left(1-a_{*}\right)^{1/3}\right]
\end{equation}
\begin{equation}
z_{2}=\left(3a_{*}^{2}+z_{1}^2\right)^{1/2}.
\end{equation}

Finally, we rewrite the equation (\ref{basic_eq1})~\&
(\ref{basic_eq2}) by using the normalized variables as
\begin{equation}
\label{solved_eq1}
\frac{d\ln{m}}{dt}=\frac{1}{\tau_{\rm EDD}}\left(\tilde{e}_{\rm in} - \eta_{\rm BZ}\right),
\end{equation}
\begin{equation}
\label{solved_eq2}
\frac{da_{*}}{dt}=\frac{1}{\tau_{\rm EDD}}\left[\left(\tilde{l}_{\rm in} -
  2a_{*}\tilde{e}_{\rm in}\right) - 2\eta_{\rm
    BZ}\left(\frac{\tilde{r}_{\rm H}}{k a_{*}} - a_{*}\right)\right],
\end{equation}
where symbols are the Eddington time $\tau_{\rm EDD}=M/\dot{M}_{\rm
  EDD}$, the specific energy input $\tilde{e}_{\rm in}=e_{\rm
  in}/c^{2}$, the efficiency of the Blandford-Znajek mechanism
$\eta_{\rm BZ}={\cal P}_{\rm BZ}/\dot{M}_{\rm EDD}c^{2}$, the specific
angular momentum input $\tilde{l}_{\rm in}=cl_{\rm in}/GM$, the
horizon radius $\tilde{r}_{\rm H}=c^{2}r_{\rm H}/GM=1 +
\left(1-a_{*}^{2}\right)^{1/2}$, and $k=\Omega_{\rm F}/\Omega_{\rm
  H}=1/2$ for the maximum efficiency of the Blandford~\& Znajek
mechanism.  Here we assume that the inner boundary is at the
marginally stable circular orbit and both the energy and the angular
momentum of accreting matter at the boundary are advected into the
black hole.  The specific energy and the specific angular momentum at
the boundary are:
\begin{equation}
\tilde{e}_{\rm in}=\tilde{e}_{\rm ms}=\sqrt{1 - \frac{2}{3\tilde{r}_{\rm ms}}},
\end{equation}
\begin{equation}
\tilde{l}_{\rm in}=\tilde{l}_{\rm ms}=2\sqrt{3}\left(1 - \frac{2a_{*}}{3\sqrt{\tilde{r}_{\rm ms}}}\right).
\end{equation}
We numerically integrated equations (\ref{solved_eq1})~\&
(\ref{solved_eq2}) with given initial parameters and track the
evolution of black hole mass and spin.  We also determined the
equilibrium spin for $m=10, 10^{6}, 10^{8}$ by solving $da_{*}/dt=0$
in the equation (\ref{solved_eq2}) by using bisection method.

Figure 3a shows the equilibrium value of spin as a function of
$\alpha\dot{m}$.  The equilibrium spin becomes larger when either
  $\alpha$ or $\dot{m}$ becomes larger.  The best-fit spin parameter
determined by the discseismic method corresponds to an equilibrium
value of $\dot{m}\approx 1$.
Figure 3b shows the time evolution of spin, where the spin parameter
of each model converges to a unique value regardless of the initial
one.  When the mass accretion rate is regulated by the Eddington value
($\dot{m}=1$), the spin converges to the equilibrium value $\approx
0.55$ for stellar-mass black holes within the order of $10^{8}$ years
and then slowly approaches the equilibrium value $\approx 0.4$ for
massive black holes.  When $\dot{m}=0.1$, the spin converges to a
value $\approx 0.5$ within the hubble time, but never actually reaches
the equilibrium spin.  Therefore the resultant spin is consistent with
the small unique spin $\approx 0.44$ when the mass accretion rate is
regulated by the Eddington value $\dot{m}\sim 1$ with the appropriate
disc parameters.  On the other hand, when the accretion disc is
somehow in a super-critically accreting phase, with $\dot{m}=10$, the
spin converges to the equilibrium value of $\approx 0.96$ within the
order of $10^{7}$ years. Although the equilibrium spin of the
super-critical accretion phase is larger than the unique value, it
could approach to this value during the subsequent sub-critical
accretion phase in less than $10^{9}$ years. The evolution of the
black hole mass is not affected by the initial spin parameter (see
Fig. 3c).  Note that the final mass becomes $10^{6}$ times
larger than the initial mass for $\dot{m}\ge 1$.

\begin{figure}
  \centerline{\epsfxsize=\hsize \epsfbox{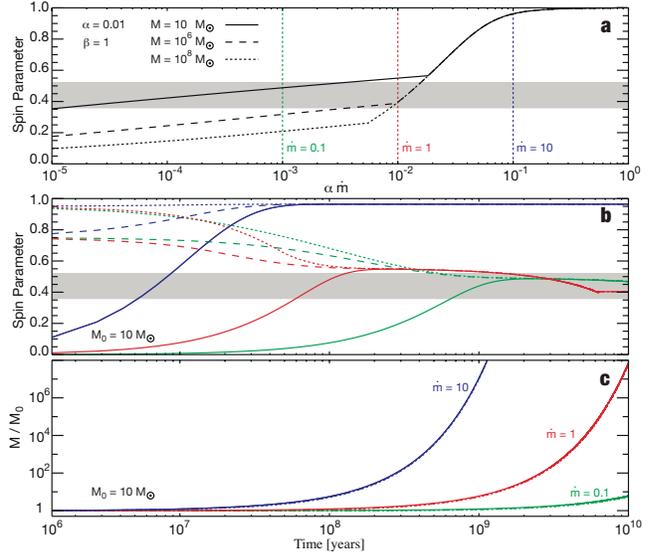}}
  \caption{Time evolution of the black hole surrounded by the standard
  accretion disc with the suitable disc parameter $\alpha=0.01$ and
  $\beta=1$ for different mass accretion rates ($\dot{m}=0.1$, $1.0$,
  and $10$ denoted by a green, red, and blue line, respectively).
  (a) is the equilibrium spin parameter in terms of $\alpha\dot{m}$
  for given black hole masses ($M=10, 10^{6}, 10^{8}M_{\odot}$ denoted
  by a black sold, dashed, and dotted line, respectively).  (b) is the
  time evolution of spin parameter for the initial black hole mass
  $M_{0}=10M_{\odot}$  with different initial spin parameters
  ($a_{*}=0.0$, $0.75$, and $0.95$ denoted by a solid, dashed, and
  dotted line, respectively).  A  gray shaded region in (a) and (b)
  indicates the best-fit spin parameter $a_{*}=0.44\pm 0.08$
  determined by the discseismic measurement.  (c) is the time
  evolution of mass ratio $M/M_{0}$ with different mass accretion rate
  and initial spin parameters.  The curves are almost independent of
  the initial spin parameters.}
  \label{fig3:eps}
\end{figure}

\section{Conclusions}
It has been suggested that the supermassive black hole in the Galactic
centre used to be in the nearly critical mass accretion phase for more
than the order of $10^{8}$ years.  A possible explanation for such
a large mass accretion history is the massive star formations in the
proximity of the Galactic centre region.  During the critical
accretion phase, the spin reaches the unique value and the mass
becomes $\sim 10^{6}M_{\odot}$, which is then maintained during the
subsequent low accretion rate phase.  Note that stochastic mass
accretion history may also help to create the moderately spinning
massive black hole (King~\& Pringle 2006).  Similarly, black holes in
Galactic X-ray sources have been in the nearly critical accretion rate
phase for order $10^{8}$ years as well, suggesting their companion
stars should be low-mass stars. Because they have reached the
quasi-equilibrium state, the limit-cycle activities and also the
emergence of jets does not alter their spin evolution.  Thus, we
conclude that the spin parameter of a supermassive black hole in the
Galactic centre has a unique value of $a_{*}=0.44\pm
0.08$. Conversely, the mass of a black hole consistent with the unique
spin is $M=\left(4.2\pm 0.4\right)\times 10^{6}M_{\odot}$.

Without detecting the event horizon, we have constrained the mass and
spin of the supermassive black hole at the Galactic centre.  The
method we used here depends entirely on geodesic frequencies that are
independent of the distance and viewing angle of a black hole.  Once
the unique spin parameter of the black hole in the Galactic centre has
been confirmed by detection of the event horizon in the future
observations (e.g., Takahashi 2004), studies of QPOs in other galaxies
will open a new window to survey the growth history of massive black
holes (Markowitz et al. 2007; Gierli{\'n}ski et al. 2008).

\section*{Acknowledgments}
The authors thank Shoji Kato and Jun Fukue for valuable discussions on
disk oscillations and also on spin evolution, and Masaaki Takahashi
and Akira Tomimatsu for helpful comments on black hole spins.
Y.K. thanks Wade Naylor for improving the English in this article.
This work was supported in part by Grands-in-Aid for Scientific
Research of MEXT [21340043, HN, Young Scientists (B) 21740149, RT].








\end{document}